\documentclass[a4paper,10pt]{article}

\usepackage{a4wide}
\usepackage{graphicx,setspace}
\usepackage{bm}
\usepackage{amsmath,textcomp}
\usepackage{amssymb}
\usepackage{subfig}
\usepackage{xcolor}
\usepackage{booktabs}

\usepackage{graphicx,setspace}
\usepackage{bm}
\usepackage{amsmath,textcomp}

\begin{document}
\vspace*{-3cm}
\begin{flushleft}
{\Large
\textbf{A multilevel analysis to systemic exposure: insights from local and system-wide information}
}
\\
Y\'{e}rali Gandica$^{1,4,\ast}$,
Sophie B\'{e}reau$^{1,3}$ and
Jean-Yves Gnabo$^{1,2}$
\\
\singlespacing
\footnotesize {
{1}  CeReFiM, Universit\'e de Namur. Belgium\\
{2} naXys, Universit\'e de Namur. Belgium \\
{3} CORE, Universit\'e catholique de Louvain. Belgium \\
{4} ICTEAM, Universit\'{e} catholique de Louvain, Belgium
}
\end{flushleft}

\singlespacing

\begin{abstract}
In the aftermath of the financial crisis, the growing literature on financial networks has widely documented the predictive power of topological characteristics (e.g. degree centrality measures) to explain the systemic impact or systemic vulnerability of financial institutions.    In this work, we 
show that considering  alternative topological measures based on local 
sub-network environment improves our ability to identify systemic institutions. 
To provide empirical evidence, we apply a two-step procedure. 
First, we recover network communities (i.e. close-peer environment) on a spillover network of financial institutions.
Second, we regress alternative measures
of vulnerability on three levels of topological measures: the global level (i.e. firm topological characteristics computed over the whole system), local level (i.e. firm topological characteristics computed over the community) and aggregated level by averaging individual characteristics over the community.  
The sample includes   $46$ financial institutions (banks, broker-dealers, insurance and real-estate companies)  listed in the Standard \& Poor's 500 index. Our results confirm the informational content of topological metrics based on close-peer environment. Such information is different from the one embeds in traditional system wide topological metrics and is proved to be 
 predictor of distress for financial institutions  in time of crisis.
\end{abstract}

\section{Introduction}

In the aftermath of the 2007-2009 financial crisis characterized by isolated shocks to   specific financial institutions being widespread to the entire system - with the Lehman brothers default in Autumn 2008 standing out as the most prominent example of such a dramatic sequence -, it has become crucial to better understand how contagion episodes   operate. Since then, increasing attention has been devoted   to understand and measure financial risk at the system level - so called ``systemic risk" - with the aim to guide policy makers in designing adequate macro-prudential regulation instruments and detect  systemically important financial institutions, i.e. the so-called  ``SIFIs". The shift of paradigm from a micro to the macroprudential perspective has contributed to approach the financial industry as an emergent system and to use tools from network science to uncover its properties. In this framework, financial institutions are assimilated to the nodes of the network and the connections between institutions are represented by the edge. The later can be financially motivated as is the case for cross lending between banks or simply the outcome of unknown mechanisms resulting in statistical dependence across institutions' stock returns. Based on this stylized
representation, a common exercise consists in applying centrality measures to flag risky institutions. The convention is to
consider, as risky for the system, institutions with several outgoing links and as fragile institutions those with several incoming
links. The analysis has been extended to other centrality measures such as Bonacich centrality, Katz centrality or Betweenness centrality (see for instance \cite{TEMIZSOY2017346} ).

While substantial efforts  in empirical financial networks have been devoted to retrieve unobserved connections, less has been done on signal extraction once the network is formed. Our claim is that centrality measures commonly used in the literature and applied at the network level  ignore important information embedded in sub-part of the network. Such information appears once the whole system is segmented in   smaller-scale and more local environments constituted of more densely connected nodes. These environments often called \textit{communities} and emanating from   privileged relationship among certain financial institutions or cross holdings for instance, have the ability to alleviate the transmission of shocks through feedback effects, being of utmost importance for risk assessment. To the best of our knowledge however, no existing study is using centrality measures by combining information at the intra  community level along with the usual system wide level. We propose to empirically explore this issue in the present contribution. 
 For this endeavor, we use the market-based approach developed in \cite{Geraci}  based on Time-Varying Parameter Vector AutoRegressive (TVP-VAR) model as well as Granger causality statistical tests on stock market returns 
to recover the unobserved spillover network of financial institutions. From there, we use the Louvain algorithm to identify the different communities constituting the network. This step is crucial to construct a new set of centrality metrics exclusively based on close-peer information. Next, we compute topological characteristics at three different levels:  (i)  the global level (i.e. firm topological characteristics computed over the whole system), (ii) local level (i.e. firm topological characteristics computed over the community), and (iii) aggregated level, by averaging individual characteristics over the community. We refer to them respectively as Global topological metrics (GT), Local topological metrics (LT) and Aggregated topological metrics (AT).   To assess the informational content of each measure, we apply Elastic Net regression techniques and test whether they can explain bank losses during the financial crisis.  
  Our sample includes 46 financial institutions embedding banks, broker-dealers, insurance and real-estate companies listed in the Standard \& Poor's 500 index. A evoked down the line, it is similar in size or larger than most related studies on market-based financial spillovers. 
  
  \section*{Literature review}
 
Our paper relates to the fast growing literature on contagion in financial network (see \cite{Glasserman} for a survey). A central question in this body of research is whether the network structure enhances the vulnerability of  the financial systems as a whole and of  its components.  A common  measure to characterize the network structure is the degree distribution. Boss et al. \cite{Boss}, for instance, document this question by using real data on interbank liabilities  in Austria. A similar exercise is proposed in \cite{Santos} and \cite{Moussa} for the Brazilian interbank network. In \cite{Caldarelli}, Caldarelli et al. argue  that, like in other types of networks such as social media, degree distribution in financial networks are found to be well described by a power law. An interesting feature emerging from this research is the identification of  core\textendash  periphery structure in financial market especially in  interbank networks. Thereby, the system is constituted of a small number of highly interconnected banks at the core along with poorly connected institutions  in the periphery (see \cite{CRAIG2014322}). Part of the literature  shifts the perspective from the system to individual nodes and extends the analysis to centrality measures which are designed to identify the importance of a node in a network (\cite{Newman_10}). One of the primary target of this research is then to assess whether the centrality of nodes explains their financial vulnerability or   systemic importance. To document this issue, several centrality measures coming from the network science have been applied to financial systems such as degree centrality, eigenvector centrality  or Katz centrality to cite the most frequent. For instance, in \cite{Craig2}, Craig et al. apply a modified version of the    eigenvector centrality on German credit register to explain individual bank risk. They find a negative relationship between   centrality measures and the probability of default. In \cite{MARTINEZ}, Martinez-Jaramillo et al, compute several centrality measures into a composite measure on the Mexican data to characterize the banking system. In \cite{Puhr}, the authors find a positive relationship between the  Katz centrality and systemic risk.  In these examples, centrality measures are applied on single layer networks assuming implicitly a unique source of connection between financial institutions such as contractual obligation. The reality is however more complex. Potential transmission channels are divers (e.g. cross-lending, derivatives, similarity, common exposure) leading to multiple corresponding network structures. 
 In order to feature the different layers, one strand of the literature has developed a holistic approach   that considers multiple channels of transmission while keeping the network representation simple. To this end, the links are recovered from  dependence in stock market price. The approach builds on the premise that stock price reflects all relevant information regarding an institution.  As such the dependence between stock returns enables to assess whether two institutions are related, regardless of the specific channels through which the transmission occurs. Thereby, it provides a synthetic measure of interconnectedness between institutions. A pioneer contribution in this vein has been done by Billio et al. \cite{Billio_2012}, where the authors use Granger-causality test, over monthly returns of hedge funds, banks, broker/dealers, and insurance companies to build their network. The authors find that it is possible to predict systemic risk of financial institutions out-of-sample with centrality measures.
Further studies in the same spirit were performed in \cite{DieboldYilmaz2016,Giudici,Hautsch}. In a previous study (\cite{plosone2018}), the presence of sub-structures within the network of densely connected nodes have been documented . Using community detection algorithm, sub-networks across time in the US market were identified. An interesting finding from this piece of research is that these sub-network stemming from   stock returns dependence does not fully coincide with trivial ex-ante categories such as financial industries.  

Our contribution relates closely to the analysis developed in \cite{Billio_2012}. The main differences are threefold. First, we apply a modified procedure, as developed in  \cite{Geraci} to retrieve the financial network. As discussed in \cite{Geraci}, this approach is better suited when the underlying network is time varying. Second, we segment our network into sub-networks by applying community detection algorithm and compute topological measures at three levels:  (i) the firm level over the whole system as done in \cite{Billio_2012}, (ii) the firm level over  the community, and (iii) the community level by averaging individual characteristics.  Another related contribution is the study proposed by \cite{TEMIZSOY2017346}, where the authors examine the exploratory power of a large set of centrality measures on interbank spread.%

\section*{Dataset} 
 
The construction of a financial network is the cornerstone of our analysis. In this study, we use the financial spillover network proposed by \cite{Geraci}. The network embeds the  financial institutions with Standard Industrial Classification (SIC) codes from 6000 to 6799 from the S\&P 500.   In addition,  we collect the monthly cum-dividend stock price from Thomson Reuters Eikon for each company. The initial database, as in \cite{Geraci}, covers the period January 1990 to December 2014 and consists of $155$ firms. Three subsequent filters on this initial   network were needed in order to recover the final network on which pre-crisis topological measures are computed. First, we remove institutions appearing in the sample only after the burst of the financial crisis in January 2008. 
Second,    we restrict our analysis to stocks with at least $36$ consecutive monthly observations in order to  filter out the noise in our data. Eventually, we drop the financial institutions that disappear during the financial crisis. Our final sample consists of  $46$  financial institutions embedding banks, broker-dealers, insurance and real-estate companies listed in the Standard \& Poors 500 index. Note that such a network is similar in size or larger than most related studies on market-based financial spillovers such as \cite{korobilis2018}, \cite{Wang} or \cite{Balla_et_al_2014}. 



\begin{figure}[!h]
\centering
\includegraphics[width = 15cm]{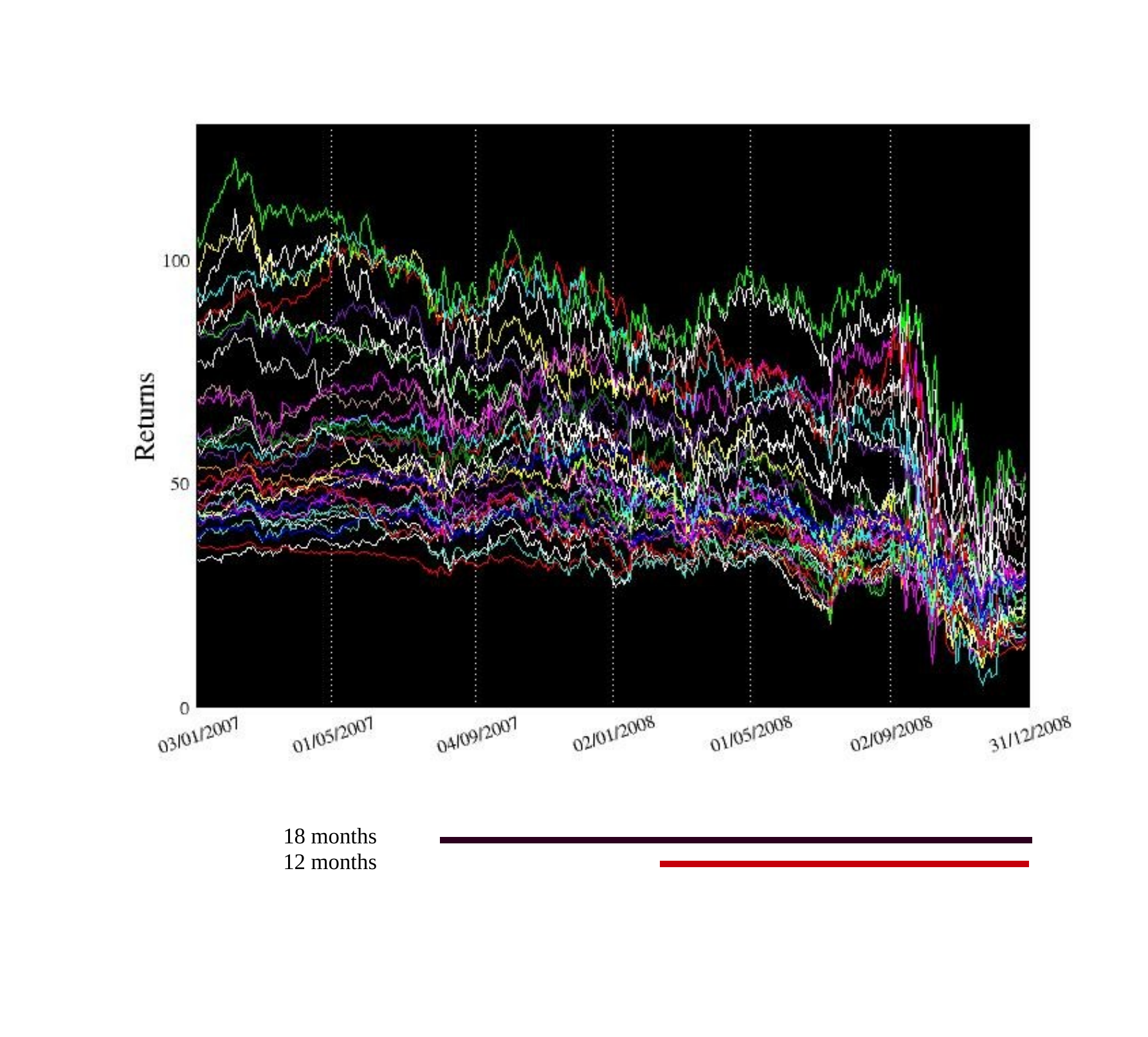}
\caption{{\bf Stock market returns  of our sample of US financial institutions (2007-2008).  }}
\label{fig1}
\end{figure}

\section*{Methodology}
First, we recover the financial network for the set of institutions comprised in our sample. Second, we break down the whole network in sub-networks, based on  a community detection algorithm and construct  topological measures at system wide and community levels.  Third, we compute a measure of vulnerability during the crisis.    Fourth, we regress our measure of vulnerability on standard topological measures along with community-based topological measures to assess their marginal contribution. Each step is detailed bellow. 
\subsection*{Step 1: Financial Temporal Networks}



  Our financial networks were built as in \cite{Geraci}. The linkages between financial institutions are retrieved from stock market prices by means of Bayesian Time-Varying VAR framework and Granger causality test. According to this approach,   an incoming link is created from institution $j$ to institution $i$ if the time series associated to $j$  Granger causes  the time series of $i$. 
The details of the methodology can be found in   \cite{Geraci}. 

\subsection*{Step 2: Independent variables}

\begin{figure}[!h]
\centering
\includegraphics[width = 10cm]{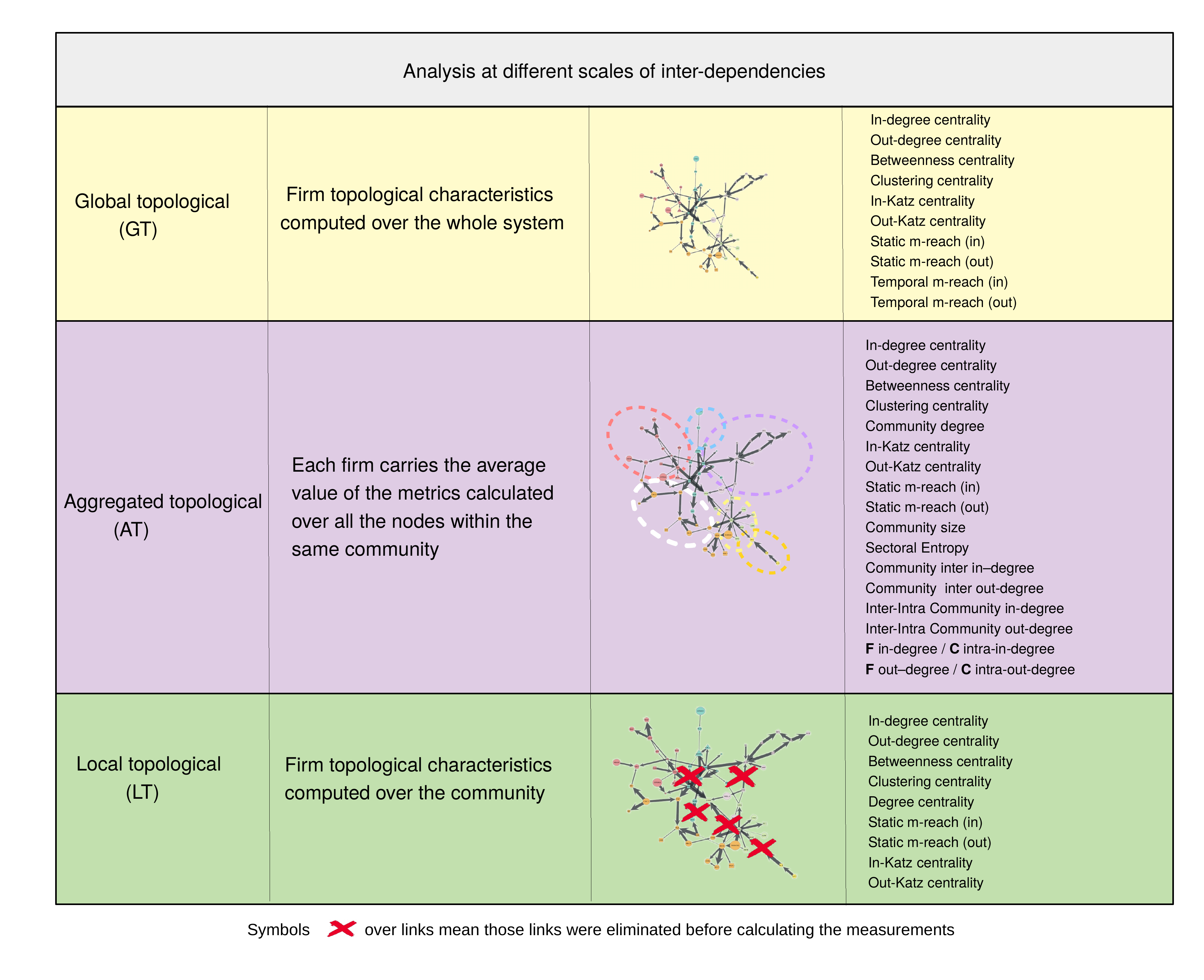}
\caption{{\bf Three levels of analysis:   System wide firms-based metrics, Community firm-based metrics and Community-based metrics.}}
\label{fig2}
\end{figure}

In a second step, we break down the whole network in sub-networks grouping  together most connected nodes. To this end, we apply  the Louvain method \cite{Blondel2008} which is designed to detect communities within complex networks. 
Equipped with clearly identified community structures, we compute several topological measures by considering the whole network or only the community-based local environment. For the sake of clarity, the whole set of variables is   divided into three groups.  First,
\textit{global topological} metrics are computed at the firm level over the whole system as usually done in the literature. Second, we compute the  \textit{local topological} metrics at the firm level over close peers. Unlike the previous metrics, for each node, we are exclusively considering the other nodes within the same community when computing our metrics, making topological metrics limited to intra-community links.    Third, in order to further    explore the influence of the local environment, we average individual characteristics over each community to compute \textit{aggregated topological} metrics. This means that all the nodes included in the same community display similar values for this set of variables. Turning to the  topological metrics we are considering, we account for a wide set of standard metrics: \textit{In-degree centrality}, \textit{Out-degree centrality}, \textit{Betweenness centrality}, \textit{Clustering centrality},\footnote{For unweighted graphs, the clustering of a node is the fraction of possible triangles associated to that node.}  \textit{m-reach centrality}, \textit{Inverse m-reach centrality}, \textit{in-Katz centrality} and \textit{Out-Katz centrality} (detailed explanation of each  variable can be found in \cite{plosone2018}). To take advantage of historical changes in the network as well as auxiliary information on firms' industry, we complete our analysis with a set of less conventional metrics. Hence,  sectoral entropy assesses how divers in term of sectors a community is.   Inter in(out)- degree depicts the number of in(out)-degree between communities. The inter-intra degree is computed as the ratio of the inter community degrees over the number of intra community degrees. The measure is applied separately to in degree and to out degree.   Besides,  we compute the rate between a firm's in(out)-degree and its community in(out)-degree as a measure of the node's commitment \cite{barabasi2007}.
Finally, two temporal metrics are added. For this, we simply compute a modified m-reach measure while considering that the contagious process involves a time lag. 
For instance, if   a firm displays 3 out degrees at time $t$, the value for its 1-reach centrality is 3. For the 2-reach centrality measure, we consider the connections of the 3 neighboring nodes at $t+1$ instead of time $t$ in the usual case. If this value is equal to 4,   the 2-reach centrality measure is   7. The measure takes into account a 1-period  delay at each order of the propagation process.   We consider a separate measure for both incoming links and outgoing links. Figure \ref{fig2} provides an overview of the  variables constructed for the empirical analysis. \\

\subsection*{Step 3. Dependent variables} 
Following \cite{Balla_et_al_2014}, we use two  indicators of firm vulnerability to test the predictive power of our topological measures: (i) the cumulative stock returns, and alternatively (ii) the peak-to-trough returns. The former measure is based on daily log-returns. The latter on prices. The cumulative returns is computed for each firm, as the sum of its stock returns over the crisis period. The maximum drawdown is computed as the maximum loss from a peak to a trough of the stock price. As done in the literature, we build on the premise that firms vulnerability is better revealed in crisis period. The crisis is therefore used as a natural experiment to identify vulnerable institutions and test whether such fragility could have been detected prior the event by looking at alternative characteristics such as topological metrics. For  robustness purposes, we use two definitions for the crisis period: (i) July 2007 to December 2008 and January 2008 to December 2008.   Figure \ref{fig1} illustrates the general patterns for all the firms. The alternative time periods to define the crisis are marked with a straight line at the bottom of the figure. A rapid overview shows that financial firms experienced largest losses during the second period (January 2008 to December 2008). 

\subsection*{Step 4: Regression analysis} 
Given the large number of regressors, we  apply the Elastic Net (EN) model  to perform shrinkage and identify the most relevant variables\footnote{ Computationally computed by using Glmnet \cite{glmnet}}. This approach along with LASSO has been widely used in the literature  to control the degree of sparsity in the model and select only the most significant coefficients \cite{glmnet_teo}. The objective function in Elastic Net regression is expressed as follows: 

\begin{equation}
    min_{\beta_0,\beta} \{ \frac{1}{N} \sum_{i=1}^{N} ( y_{i} - \beta_0 - \beta^T x_{i})^2 + \lambda [(1- \alpha) ||\beta||_2^2 / 2 + \alpha ||\beta||_1] \},
\end{equation}

The EN penalty is controlled by the parameter $\alpha$. Note that if $\alpha=1$ it falls into  the LASSO selection model. The second   parameter  to tune in order to set the penalty is $\lambda$. It more specifically controls the strength of the penalty, i.e, the number of covariates set to zero (if $\alpha\neq0$) or shrink towards zero (if $ \alpha=0$) for minimization purposes. We chose the $\lambda$ values via cross validation for fixed  $\alpha$. We consider   0.5 and 1 as alternative values for $\alpha$. \\
The regression framework is convenient to assess the role of local environment information as opposed to traditional global environment information for computing topological metrics and analyzing systemic risk. For the sake of clarity, we can more specifically formulate three testable hypotheses. \\ 
\begin{itemize}
     \item \textbf{H1}: Information regarding systemic vulnerability of firms  is concentrated at the \textbf{local level}. If so,  information coming from the remaining part of the network only  add noise and should be filtered out.  
\item \textbf{H2}: Information regarding systemic vulnerability of firms in the local environment is entirely embodied in the \textbf{global environment}.  
\item \textbf{H3}: Both \textbf{local and global levels} embeds non overlapping information. 
\end{itemize}
If \textbf{H1} is true, only LT metrics should appear as significant from a regression including both local and global topological metrics. If \textbf{H2} is true conversely, only GT should emerge as significant. If \textbf{H3} is true, both LT and GT variables should appear as significant when included in the same model.

\section*{Results}
Before interpreting the regression results, we   briefly discuss the cross correlation among the topological metrics. Exploring the dependence among all our metrics might help to have a better sense of their informational content. A very strong correlation  for instance would suggest the absence of specific information in each variable.   Next, we estimate a baseline model. This model is limited to traditional centrality metrics computed at the network level. Then, we extend the specification to include community-based metrics.  
Eventually, we include all the variables in the model. This stepwise procedure aims to assess the robustness of our findings. 
\subsection*{The correlation matrix}
\begin{figure}[!h]
\centering
\includegraphics[width = 13cm]{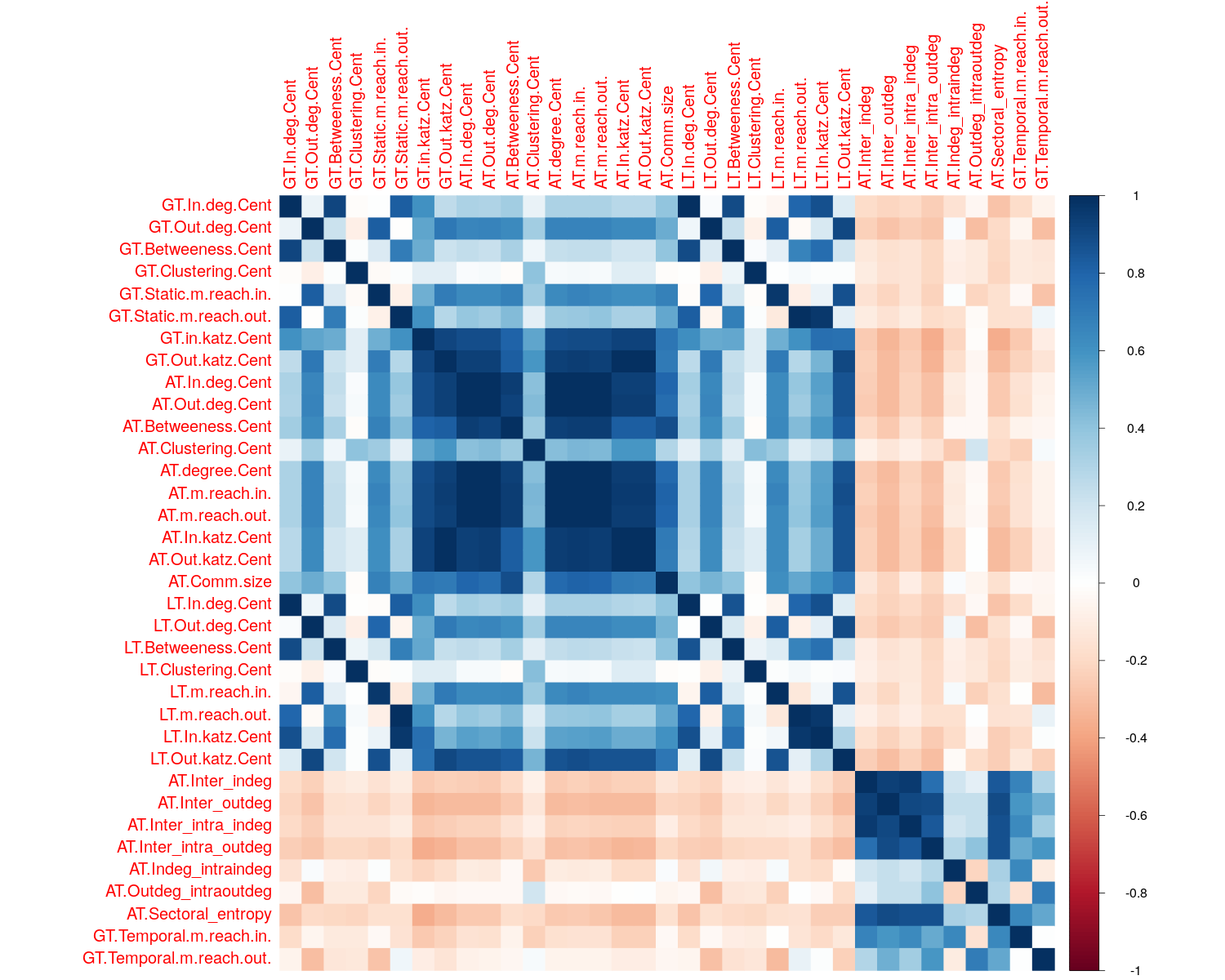}
\caption{{\bf Correlation Matrix of topological metrics.}}
\label{fig3}
\end{figure}

 For the sake of clarity, the correlation coefficients are visualized with a heatmap (see Figure \ref{fig3}). Dark colors signal strong correlations. Positive correlations are reported in blue and negative ones in red. We can visually separate two groups of variables. The first group embeds most   traditional topological metrics computed either at the network or the community level. 
 Overall, the variables display positive correlations with nuances  regarding their  strength as the correlation coefficients range from 0.1 to 0.9. The positive signs suggest that the various measures embed consistent   information regarding the centrality of institutions. While consistent however,  this information does not fully overlap. A second group includes less conventional measures such as sectoral entropy or temporal m-reach centrality measures. The correlation values are now weak and slightly negative with the majority of other variables. Here also, the result supports the interest of considering a wide variety of centrality indicators as they do not share the same information. Whether they are informative to predict financial vulnerability remains  an open question that the regressions enable to address. 

\subsection*{First level: Global topological metrics}
We first show the results for the regular metrics. In total, the model includes $8$  variables, all computed at the system-wide level. The list of variables can be found in Figure \ref{fig2}. We just not consider the two temporal (GT) variables in this baseline model.  Table $1$ reports in row the name along with the sign of the coefficient of the variables that appear as significant in one of the four columns.  Each column corresponds to an alternative definition of the dependent variable. We consider four alternatives depending on the definition of both the crisis period - either a period of 12 months or 18 months- and the vulnerability of the firm - computed either as  cumulative returns or maximum drawdown.   Four variables are identified as significant regardless of  how the dependent variable has been computed at the exception of the \textit{out degree} variable which is not significant in one model. As expected, the signs for the cumulative returns and the maximum drawdown are opposite. Higher vulnerability corresponds to   smaller cumulative returns and   larger maximum drawdown. To illustrate our results, we can take the example of  the \textit{Betweenness centrality} metric. Remind that \textit{Betweenness centrality} is the number of  shortest paths that pass through a node. In our context, each edge represents the existence of spillovers between the asset returns of two financial institutions. The path means that a distress in one institutions can be channelled to another one in the network through indirect connections.   The \textit{Betweenness centrality} metric therefore captures whether a financial firm is located in the path of these financial spillovers. It can also be viewed as  a measure of the capability of a firm to transmit the crisis between separated groups of firms. Our results support that the higher the number of paths passing through a firm the greater the losses experienced by this firm during the crisis  that is the most vulnerable the firm is.

\begin{table}[!h]
\label{tab1}
\begin{tabular}{l l l l l}
\toprule
\textbf{Significant variables} & \textbf{C. R:12 m} & \textbf{C. R:18 m}  & \textbf{M. D:12 m}  & \textbf{M. D:18 m}\\
\midrule
Out-degree centrality (GT)&$+$ & $+$ &  &$-$  \\
Betweenness centrality (GT)& $-$ & $-$ & $+$ & $+$ \\
Clustering (GT)& $-$ & $-$ & $+$ & $+$ \\
In-Katz centrality (GT) &$-$ &$-$ &$+$ &$+$\\
\bottomrule
\label{tab:1}\\
\end{tabular}
\caption{Linear multivariate regressions for system wide firm-based metrics}
{\footnotesize Note: -, + denote the sign of the coefficient for significant variable. The variables not significant are not reported. The model is estimated by using Elastic Net regression. Each column corresponds to a definition of the dependent variable. In columns 1 and 2, the dependent variable is computed as cumulative returns over respectively 12 months and   18 months. In columns 3 and 4,  the dependent variable is computed as  the maximum drawdown over 12 months and 18 months,  respectively.}
\end{table}

\subsection*{Second level: Global and   aggregated topological metrics }
In the second estimation, we add to the baseline model (GT) all the variables averaged by community (AT). By doing so, we can assess the stability of the firm-based results along with the community-level of analysis. Results are depicted in Table $2$. The traditional metrics appear robust, when compared to the first-level estimations. Only three changes  are noticeable. The \textit{outdegree} variable is now significant with respect to the four alternative definition of the dependent variable. It appeared as significant in three models previously. The \textit{Betweenness centrality} is no more significant   when we consider a long period of 18 months for the crisis and the maximum drawdown to quantify firm losses.  Eventually, \textit{in-Katz centrality} is no more significant when using a 18 month crisis period and the cumulative returns.  Among the 10 variables computed at the community level, three of them are significant once controlling for traditional system-wide metrics: \textit{clustering}, \textit{2-reach} and \textit{community size}. This result means, for instance, that the larger a community the more vulnerable a firm is. 
As opposed to clustering and size, the 2-reach centrality is a clear measure of temporal propagation and node importance.  Its significance at this level of the analysis stresses its interest   among the set of  network science metrics for systemic risk analysis.

\begin{table}[!h]
\label{tab:2}
\begin{tabular}{l l l l l}
\toprule
\textbf{Significant variables} & \textbf{C. R:12 m} & \textbf{C. R:18 m}  & \textbf{M. D:12 m}  & \textbf{M. D:18 m}\\
\midrule
Out-degree centrality (GT) & $+$ & $+$ & $-$ & $-$ \\
Betweenness centrality (GT)  & $-$ & $-$ &$+$  &  \\
Clustering centrality (GT) & $-$ & $-$ &$+$  &$+$  \\
In-Katz centrality (GT)  &$-$ &  &$+$ &$+$  \\
Clustering (AT) & $-$ & $-$ &$+$  &  \\
2-reach centrality (AT) & $-$ & $-$ &  &  \\
Community Size (AT) &  $-$ &  &$+$  &$+$  \\
\bottomrule

\end{tabular}
\caption{Linear multivariate regressions for system wide firm-based (GT) metrics and community-based metrics (AT)}
{\footnotesize Note: See Table $1$}
\end{table}

\subsection*{Third level: Complete model}
We  complete the analysis by considering the full model. The specification includes the three categories of variables, adding community firm-based metrics (noted LT in Figure 3) to the previous model (noted GT and AT in Figure 3). We also add our set of alternative metrics (noted  new-GT and new-AT in Figure 3). As explained in the Step 2 of the methodology, these variables exploit   information embeds in the temporal changes in the network, the sectoral diversity of our financial firms, as well as, the intra and extra community links. Table 3 reports the results. System-wide metrics are  stable for \textit{Betweenness} and \textit{Clustering}. Out-Katz centrality performs better than the  outdegree variable to account for the influence, adding up the influentially of the neighbouring nodes. On the other hand, inverted 2-reach centrality took the place of the in-Katz centrality to express  financial vulnerability. The size of the community remains significant, as expected and explained in the last section. Turning to the community firm-based (LT) metrics, two variables expressing out-going and in-coming  links are the dominant ones at this level. The first one, is the out-degree centrality. Interestingly, we have earlier noticed that this variable was significant when computed at the network level.  Now that the two variables computed at the network and the community level are competing in the same model the one at the community level emerges as the one significant. The second variable is the inverted 2-reach centrality. We can note that in this case the same metrics computed at both the system wide and the community level are significant while included in the same model. Among the set of alternative metrics five variables appear as significant: Inter-intra-outdegree centrality, outreg-intra-outdeg, sectoral entropy, temporal inverted 2-reach, temporal 2-reach. Recall that  \textit{Inter-intra-outdegree centrality}, is defined as the ratio between the community out-degree to the other communities in the network and the out-degree within the same community. What we want to express by such a metric is the influence that one community has over the whole system, normalized by the own intra-dependence. The higher the variable the more vulnerable the firm.  The second metrics is the outdeg-intra-outdeg. The creation of this metric was inspired by \cite{barabasi2007}, when defined as the node's commitment with its own community. Here the metric is defined as the ratio between the node's outdegree and the total outdegree within its community. The sign associated to the first column when the dependent variable is constructed as cumulative returns   is positive. The sign is negative when considering the maximum drawdown. This result means that the higher the variable the less vulnerable the firm.   The \textit{sectoral entropy} captures the diversity of sectors within community. Sectoral diversity was suggested to be a determinant systemic risk in the descriptive analysis of \cite{plosone2018}, when computed as the global sector-interface.  Our estimations  based on a formal regression analysis confirm this feature.
 It is worth noting, however, that  the metric is sensitive to the time length, appearing as significant only for the shortest crisis period ($12$ months).  Finally, we have the temporal versions of the m-reach centrality. Only the one considering outgoing links (Temporal 2-reach) is dominant. This last result point out to the importance of considering higher-order metrics into systemic risk \cite{lambiotte2019}. 

\begin{table}
\label{tab:3}
\begin{tabular}{l l l l l}
\toprule
\textbf{Significant variables} & \textbf{C. R:12 m} & \textbf{C. R:18 m}  & \textbf{M. D:12 m}  & \textbf{M. D:18 m}\\
\midrule
Betweenness centrality (GT)  & $-$ & $-$ & $+$ & $+$ \\
Clustering (GT) & $-$ & $-$ & $+$ & $+$ \\
Inverted 2-reach centrality (GT) & $+$ &$+$ & $-$ & $-$ \\
Out-Katz centralitiy (GT) & $-$ & $-$ & $+$ & $+ $ \\
Community size (AT)  & $ $ & $ $ & $+$ & $+ $ \\
Out-degree centrality (LT) & $+$ & $+$ & $-$ & $-$ \\
Inverted 2-reach centrality (LT) & $+$ & $+$ & $-$ &  \\
Inter-intra-outdeg (new-AT) & $-$ &  & $+$ &  \\
Outdeg-intra-outdeg (new GT) &$+$& $+$ &  & $-$ \\
Sectoral-Entropy (new-AT) &$+$&  &$-$ & \\
Temporal inverted 2-reach (new GT)  &  &  & & $+$ \\
Temporal 2-reach (new GT) & $-$ & $-$ & $+$ & $+$ \\
\bottomrule
\end{tabular}
\caption{Linear multivariate regressions for system wide firm-based metrics, community-based metrics and community firm-based metrics}
{\footnotesize Note: See Table $1$}
\end{table}

\section*{Conclusions}

In this study, we show that several topological characteristics of financial institutions stemming from the whole network
and sub-parts can predict their vulnerability in crisis period. More specifically, we document the fact that local information along with global information is useful to improve our knowledge of systemic institutions.    This result is obtained by applying a two-step procedure in the
spirit of \cite{Geraci}. In a first step, we use the approach developed by \cite{Geraci} based on Time-Varying
Parameter Vector AutoRegressive (TVP-VAR) model as well as Granger causality statistical tests on stock market returns
to recover the unobserved spillover network of financial institutions. In a second step, we regress alternative measures
of institutions vulnerability on a set of pre-crisis topological characteristics. We test for the marginal impact of topological
measures computed at firms level with respect to the whole network and with respect to  close peers (i.e. we only consider intra-community links), and averaged by community.   Overall, our sample include  46  financial institutions embedding banks,
broker-dealers, insurance and real-estate companies listed in the Standard $\&$ Poor's 500 index.


\begin{thebibliography}{100}
\bibitem{TEMIZSOY2017346} Temizsoy, A.,  Iori, G. and Montes-Rojas, G.  Network centrality and funding rates in the e-MID interbank market.   Journal of Financial Stability.\textbf{33}. https://doi.org/10.1016/j.jfs.2016.11.003. 2017.
\bibitem{Geraci} Geraci, M. and Gnabo, J-Y. Measuring interconnectedness between financial institutions with bayesian time-varying vector autoregressions. Journal of financial and quantitative analysis. \textbf{53} (3). 2018. 
\bibitem{Glasserman} Glasserman, P.  and Young, H. P.  Contagion in Financial Networks. Journal of Economic Literature. 54. N 3. DOI = {10.1257/jel.20151228} 2016. 
\bibitem{Boss}  Boss, M., Elsinger, H., Summer, M. and  Thurner, S.  Network topology of the interbank market. Quantitative Finance. \textbf{4}, n 6. doi:10.1080/14697680400020325. 2004.
\bibitem{Santos} Santos, E. and Cont, R.  The Brazilian Interbank Network Structure and Systemic Risk. Central Bank of Brazil, Research Department. Working Papers Series. 219. Https://EconPapers.repec.org/RePEc:bcb:wpaper:219. 2010.
\bibitem{Moussa} Cont, R., Moussa, A. and Santos, E. Network structure and systemic risk in banking systems.  Handbook of Systemic Risk. JP Fouque \& J Langsam. Cambridge Univ Press. PAGES 327-368. https://hal.archives-ouvertes.fr/hal-00912018. 2013. 
\bibitem{Caldarelli}  Caldarelli, G.,   Battiston, S.,  Garlaschelli, D. and   Catanzaro, M. Emergence of Complexity in Financial Networks.  In: Ben-Naim E., Frauenfelder H., Toroczkai Z. (eds) Complex Networks. Lecture Notes in Physics, vol \textbf{650}. Springer, Berlin, Heidelberg. https://doi.org/10.1007/978-3-540-44485-5\texttildelow18. 2004.
\bibitem{CRAIG2014322}   Craig, B.  and  von Peter, G.  Interbank tiering and money center banks. Journal of Financial Intermediation. \textbf{23}, n 3, https://doi.org/10.1016/j.jfi.2014.02.003. 2014. 
\bibitem{Newman_10} Newman, M. Networks: an introduction. Oxford University Press. 2010.
\bibitem{Craig2} Craig, B., Koetter, M. and Kruger, U. Interbank lending and distress: Observables, unobservables, and network structure.  Deutsche Bundesbank. Discussion Papers. 18/2014. https://ideas.repec.org/p/zbw/bubdps/182014.html. 2014. 
\bibitem{MARTINEZ}  Martinez-Jaramillo, S., Alexandrova-Kabadjova, B.,   Bravo-Benitez, B. and  Sol\'{o}rzano-Margain, J.P.  An empirical study of the Mexican banking systems network and its implications for systemic risk. Journal of Economic Dynamics and Control. \textbf{40}. https://doi.org/10.1016/j.jedc.2014.01.009. 2014. 
\bibitem{Puhr}  Puhr, C., Reinhardt, S. and  Sigmund, M. Contagiousness and Vulnerability in the Austrian Interbank Market.  Financial Stability Report. \textbf{24}. 2012. 
\bibitem{Billio_2012}  Billio, M.,  Getmansky, M. and  Lo, A. W. and  Pelizzon, L.  Econometric measures of connectedness and systemic risk in the finance and insurance sectors.           Journal of Financial Economics. \textbf{104}. n 3. 2012. 
\bibitem{DieboldYilmaz2016} Diebold, F. X. and Yilmaz, K.  On the Network Topology of Variance Decompositions: Measuring the Connectedness of Financial Firms.  Journal of Econometrics. \textbf{182}. 2014. 
\bibitem{Giudici} Giudici, P. and Parisi, L. Risks. Bail-In or Bail-Out? Correlation Networks to Measure the Systemic Implications of Bank Resolution.  \textbf{7}. n 1. https://www.mdpi.com/2227$-$9091/7/1/3. 2019. 
\bibitem{Hautsch}  Betz, F.,  Hautsch, N.,    Peltonen, T. A. and  Schienle, M. Systemic risk spillovers in the European banking and sovereign network.  Journal of Financial Stability. \textbf{25}. https://doi.org/10.1016/j.jfs.2015.10.006. 2016.
\bibitem{korobilis2018} Korobilis, D. and Yilmaz, K.  Measuring Dynamic Connectedness with Large Bayesian {VAR} Models. Essex Finance Centre Working Papers. 2018. 
\bibitem{Wang} Wang, D.,  van Lelyveld, I. and  Schaumburg, J. Do information contagion and business model similarities explain bank credit risk commonalities? Tinbergen Institute. Tinbergen Institute Discussion Papers. 18-100/IV. 2018. 
\bibitem{Balla_et_al_2014}  Balla, E.,  Ergen, I. and   Migueis, M.  Tail dependence and indicators of systemic risk for large {US} depositories. Journal of Financial Stability. \textbf{15}. 2014. 
\bibitem{Blondel2008} Blondel, V., Guillaume, J-L, Lambiotte, R. and Lefebvre, E. Fast unfolding of communities in large networks.   Journal of Statistical Mechanics. \textbf{10}. P10008. 2008. 
\bibitem{plosone2018}  Gandica, Y., Geraci, M., B\'ereau, S. and  Gnabo, J-Y.  Fragmentation, integration and macroprudential surveillance of the US financial industry: Insights from network science. PLoS ONE \textbf{13}(4): e0195110. \\
\bibitem{glmnet} https://www.rdocumentation.org/packages/glmnet/versions/2.0-16/topics/glmnet
\bibitem{barabasi2007}  Palla, G.,   Barabasi, A-L and   Vicsek, T.  Quantifying social group evolution. Nature, \textbf{446} (2007) doi:10.1038/nature05670
\bibitem{glmnet_teo} https://web.stanford.edu/$\sim$hastie/Papers/Glmnet\texttildelow Vignette.pdf
\bibitem{lambiotte2019} Lambiotte, R., Rosvall, M.  and Scholtes, I.  From networks to optimal higher-order models of complex systems. Nature Physics. \textbf{15}, 2019. 



\end{thebibliography}

\section*{Acknowledgments}
We would like to thank Marco Geraci for sharing his database with us. We thank Renaud Lambiotte and Timoteo Carletti for interesting scientific discussions. We are also grateful to the Benet 2017 participants.

\end{document}